\documentclass[11pt, oneside]{article}   	% use "amsart" instead of "article" for AMSLaTeX format

\usepackage{geometry}                		% See geometry.pdf to learn the layout options. There are lots.
\usepackage{graphicx}				% Use pdf, png, jpg, or eps§ with pdflatex; use eps in DVI mode
\usepackage{amssymb,bm,hyperref}
\usepackage[T1]{fontenc}
\usepackage[english]{babel}

\title{1930-1937: the first $\beta$-rays and  neutrino theories}
\author{Francesco VISSANI\\
\sc\small INFN, Laboratori Nazionali del Gran Sasso}
\date{\small Padua, September 2023\footnote{Presented at: 43rd National Congress of the Italian Society for the History of Physics and Astronomy, 
Sept.~2023, Padua. 
Published as: Vissani, F., {\em 1930-1937: The First $\beta$-ray and Neutrino Theories}, 
Atti del XLIII Convegno annuale degli Storici Italiani della Fisica e dell'Astronomia (SISFA),
pp.~287- 294 (2024).}}							% Activate to display a given date or no date

\begin{document}
\maketitle
%\section{}
%\subsection{}

\begin{abstract}
The conceptual bases of Fermi's $\beta$-ray theory (at its 90th anniversary) are examined, 
highlighting the innovative drive and inspirational role for the progress that followed just afterwards.
Moreover, the three different ideas of the neutrino born from the proposals of Pauli 1930 \cite{pauli}, again 
Fermi 1933 \cite{rs33} and Majorana 1937~\cite{maj} are discussed, emphasising 
the interest of the latter for current expectations.
%\footnote{Proceedings of the 43rd Annual Conference of Italian Historian of Physics and Astronomy, 2023. 
%Cite as: Vissani, Francesco, {\em 1930-1937: The First $\beta$-ray and Neutrino Theories}, 
%Atti del XLIII Convegno annuale degli Storici Italiani della Fisica e dell'Astronomia (SISFA),
%5pp. 287- 294 (2024).}
\end{abstract}

\begin{quote}
\end{quote}

\parskip0.5ex
\section{Introduction}

On the 50th anniversary of the discovery of radioactivity
induced by neutrons, Edoardo Amaldi wrote a monumental work of
review on those very topics \cite{amuld}. This work 
contains much valuable and unique material: e.g., there is a famous footnote, in which  
the origin of the word ``neutrino'' is recounted. We do not quote its 
text, relying on the fact that this story is already known,  
and we deal instead with Amaldi's presentation of an
important and closely related aspect, only apparently technical.

In the section entitled ``Fermi's paper on beta decay'' (page 82)
there is a description that leads the modern reader to spontaneous assent, this one:
\begin{quote}
\sf \small
his density of interaction Hamiltonian
$H_{fi}$ is expressed as the product of 2 four-vectors computed at the
same point (contact interaction), one concerning the heavy particles,
the other the light
particles:
\begin{equation}
H_{fi} = g
\left[ 
\left(  \bar{\psi}_{p} \gamma_{\mu}  \psi_{n}     \right) 
\left( \bar{\psi}_{e}   \gamma^{\mu} \psi_{\nu} \right) 
+ \mathrm{h.c.} 
\right]
\end{equation}\label{frengi}
\end{quote}
However, scrolling the text of the original works, written from 1933 to 1934~\cite{rs33,nc34,zs34}, it is easy to convince oneself that Fermi uses a Hamiltonian, not a `Hamiltonian density' as in eq.~(\ref{frengi}); that his description of nucleons does not rely on the relativistic formalism (and in this way, the symmetry between hadrons and leptons is not emphasised); that neither Dirac \(\gamma_{\mu}\) matrices nor Dirac conjugates are mentioned; that Fermi talks of neutrino emission, not of antineutrinos. Eq.~\ref{frengi} is a modern expression, that in a sense corresponds to those in~\cite{rs33,nc34,zs34}, 
to which modern theoretical physicists are accustomed, but that  
does not allow us to understand the difficulties encountered and overcome by Fermi,
and that also prevents us from appreciating the value of subsequent theoretical progress.

In view of the fact that Amaldi's review paper has been (and is) influential, and presentations similar or identical to his have since been commonly adopted - see e.g., \cite{bil} - 
we propose to consider a series of questions to prepare ourselves to better appreciate Fermi's work and legacy:
\begin{itemize}\em
\item \small
What are the objectives and conceptual bases of Fermi's theory? What are its radical innovations?
\item \small
What was Fermi's theory important for at the time?
\item \small In what aspects does Fermi's theory of $\beta$ decay differ from
the modern one?
\item\small
How do Pauli's, Fermi's and Majorana's ideas on the
neutrino compare with each other?
\end{itemize}
In the following discussion,  we will draw mainly on a recent
article \cite{90} prepared on the occasion of the 90th anniversary of Fermi's  `Tentativo', 
to which we refer the reader interested in detailed information and specific references.
%the first successful theory of $\beta$ rays.
\normalsize

\section{Fermi's  theory of $\beta$ rays}

\subsection{Origin, purpose, basis and innovations}
The aim of Fermi's work is to provide an answer to the question 
{\em``how is it possible for the nucleus to emit electrons, if there are no electrons in the nucleus?''.} The formulation of this question helps us to remind the state of previous knowledge: in the second decade of the 20th century, a somewhat spontaneous 
opinion gained traction, that the electrons emitted in the $\beta$ decay must pre-exist in the nucleus. This view is clearly stated in a well-known work of 1920 by Rutherford  \cite{ruddy}, in which he adheres to a model of a nucleus consisting of protons and electrons. 
As soon as the neutron is discovered, a new and better model  is proposed, 
where the nucleus contains only protons and neutrons
\cite{n1,n2,n3} (Iwanenko 1932, Heisenberg 1932 and Majorana 1933);  
but this urgently raises the question of how to model the emission of $\beta$ rays.

Inspired by de Broglie's ideas, Ambarzumian and Iwanenko \cite{ai} had suggested already in 1930 that the electron is created in that process, just as happens to a photon spontaneously emitted by an excited atom; the same was further advocated in 1933 by Francis Perrin, for whom the neutrino should also suffer a similar fate \cite{cr}. But none of these authors succeeded in producing a quantitative theory, a calculable model.

Fermi, on the other hand, succeeded in this endeavour with the three papers mentioned above, which have the same content. The first of these appeared just 90 years ago, and the other two provide some further details. The model describes the situation in which an atomic nucleus increases its charge by one unit (attributing this to a change of state of a nucleon -  from a neutron,  to a proton) and at the same time an electron and a neutrino are created. In formulae,
\begin{equation}(A,Z) \rightarrow (A,Z + 1) + e + \nu\end{equation}
It is well-known that Fermi's description is generally in good agreement with the observational facts and, over time, it has been improved in various aspects, rather than radically modified. 

But let's take a closer look at its original structure.
The mathematical formalism adopted to deal with relativistic fermions 1)~assumes the correctness of the Dirac equation, 2)~the 
interpretation of its spectrum due to Dirac (see below), and 3)~exploits the technique of second quantisation developed by Jordan, Klein, Wigner and Fock. This formalism implies 
using operators 
\begin{equation}\mathbf{\Psi} = \sum_{s}^{}\psi_{s}\, \mathbf{a}_{s}\label{cicciolo}\end{equation}
with dimensions square root of a density; the sum is over all possible states \(s\) (positive and
negative energies); \(\psi_{s}\) are wavefunctions that solve Dirac  equation, normalized {\em \`a la} Born; 
 \(\mathbf{a}_{s}\) 
are adimensional annihilation  operators  that describe the disappearance of a particle in the state 
\(s\):
$\langle 0\left| \mathbf{a}_{s} \right|s\rangle = 1$. 

How to avoid a disastrous process of creating electrons of negative energy - and in particular, how to prevent atomic electrons from accessing negative energies?
The chosen way to go is the one described by Dirac, that we are going to recall. 
It is assumed that, as a rule, all negative energy fermion states 
are occupied; this is the hypothesis of the {\bf\em Dirac sea}. We reiterate
that this formalism is used {\em only} for electrons and
neutrinos. In this way, Fermi
%
%Di regola, lo
%stato di vuoto di ogni fermione \`e completamente occupato nella regione
%delle energie negative; questa \`e l'ipotesi del mare di Dirac. Ribadiamo
%che questo formalismo viene utilizzato \emph{solo per elettroni e
%neutrini.} In questo modo, Fermi
\begin{itemize}
\item 
manages to describe the spin of electrons and neutrinos in the theory;
%riesce a descrivere lo spin degli elettroni e dei neutrini nella teoria;
\item
does not emphasise the other crucial aspect of Dirac  equation - 
antiparticles;
%non d\`a risalto all'altro aspetto cruciale dell'equazione di Dirac, le
%antiparticelle;
\item
relies on the less innovative - but adequate - isospin formalism for
nucleons.
%si affida al meno innovativo - ma adeguato - formalismo dell'isospin per
%i nucleoni.
\end{itemize}
Let us emphasise the point  we made, as explicitly as possible:
\begin{quote} \emph{Fermi uses quantized fields to deal with electrons and neutrinos, but not the formalism of canonical quantisation}
\end{quote}

The original form of Fermi's hamiltonian is the following one,  % La forma originaria della hamiltoniana di Fermi \`e la seguente
\begin{equation}{\bm{H}} = g\, {Q}\, \bm{\Psi}^{t}\delta\bm{\Phi} + \mathrm{h.c.}\label{maronz}
\end{equation}
where \newline$\star$
$g$ denotes Fermi's constant, with units energy per volume; \newline$\star$
$ {Q}$ the dimensionless isospin matrix, which transforms a proton into a neutron; \newline$\star$
$\bm{\Psi}$ and  $\bm{\Phi}$ the fields of second quantisation of the relativistic particles with spin 1/2, the electron and the neutrino, which have the same units as the wave functions (root of  a density=root of an inverse volume);\newline$\star$
the superscript $t$ denotes the transpose; \newline$\star$
$\delta$ an appropriate $4\times 4$ dimensionless matrix which ensures the Lorentz invariance of the expression;
\newline$\star$ the Hermitian conjugate  term ($h.c.$) in Eq.~\ref{maronz} guarantees the conservation of probability.\\
The interaction energy    $\bm{H} = \int d^3x\, e V \bm{\Psi}^\dagger \bm{\Psi}$ is 
the model of Fermi's hamiltonian: 
the 
electrostatic energy $e\, V(x)$ 
is replaced by the nuclear energy 
$g\, \delta^3(x) {Q}$ and the scalar current $\bm{\Psi}^\dagger \bm{\Psi}$ by  $\bm{\Psi}^{t}\delta\bm{\Phi}$;  
the expression is given in the limit of nucleons at rest. 
The $h.c.$ term accounts for the decay of the neutron through the irradiation of an electron and a neutrino.

From a conceptual point of view, the main innovation of Fermi's model is that it formally describes the possibility that a particle can be destroyed or created. It is the first time that {\em particles of matter} are assumed to undergo a similar fate. 
This  constitutes a milestone in modern particle physics, although the formalism adopted (which derives from Jordan, Klein on the one hand and from Dirac's positron theory on the other) does not coincide with the current one.
See again~\cite{90} for further discussion and just below for subsequent developments.

%as one might be led to believe by Amaldi's presentation.

%dove \`indico con \(g\) la costante di Fermi, con unit\`a di misura
%\emph{energia per volume}; con \(\mathbf{Q}\) la matrice di isospin
%adimensionale, che trasforma un protone in un neutrone; con
%\(\mathbf{\Psi}\) e con \(\mathbf{\Phi}\) i campi di seconda
%quantizzazione delle particelle relativistiche dotate di spin 1/2,
%l'elettrone e il neutrino, che hanno le stesse unit\`a di misura delle
%funzioni d'onda (radice dell'inverso di un volume); con \(\delta\) una
%opportuna matrice adimensionale \(4 \times 4\) che assicura l'invarianza
%di Lorentz dell'espressione. L'espressione \`e data nel limite di nucleoni
%a riposo. Da un punto di vista concettuale, l'innovazione principale del
%modello di Fermi \`e quella di aver descritto matematicamente la
%possibilit\`a che una particella possa essere distrutta o creata. Si noti
%che \`e la prima volta che questo viene fatto per particelle \emph{di
%materia.}

\subsection{Reactions to Fermi's paper and its  legacy}

Fermi's use of Dirac  sea exposed him at the time to the same criticism as Dirac\footnote{See e.g., \cite{pais}  and \cite{kragh}. In his memoirs, Occhialini reiterates that the old guard physicists such as Rutherford and Bohr, but also Chadwick, maintained reservations at least until 1932. From Majorana's correspondence, a feeling of doubt towards Dirac  interpretation persisted in 1933.
% L. Brown  mentions Landau and Fock among the sceptics, and 
In 1933, Pauli still doubts Dirac's argument for the anti-electron; next year, with Weisskopf he succeeded in quantising a hypothetical spinless particle without resorting to Dirac  sea - a procedure Pauli liked to refer to as `anti-Dirac theory'. See \cite{90} for references.}.  In addition to these reservations of a general nature, the work will be the subject of a wide-ranging and lively debate. Limiting ourselves for the moment to the main contributions of a critical nature, let us mention for example the specific 1935 proposal by Konopinski and Uhlenbeck \cite{ku}, which at first seemed superior to Fermi's, but which emerges defeated from the confrontation a few years later. Let us then recall a criticism by Pauli in 1938 \cite{p38}, centered on the fact that the theory includes the parameter $g$ with canonical dimensions equal to the inverse of a square mass in natural units, a circumstance that entails going outside the theory itself with perturbative orders higher than the first; but as is well known today, Fermi's theory is to be thought of as an effective theory and therefore is to be used precisely at the first perturbative order.

In short, Fermi's theory fully hits the mark, in spite of the usage of second quantization based on Dirac sea (or in Fermi's words, the Dirac, Jordan, Klein procedure) and the specific choice of Hamiltonian function, aspects that only apparently are limiting.

\begin{figure*}[p!]
\centerline{\includegraphics[width=0.99\textwidth]{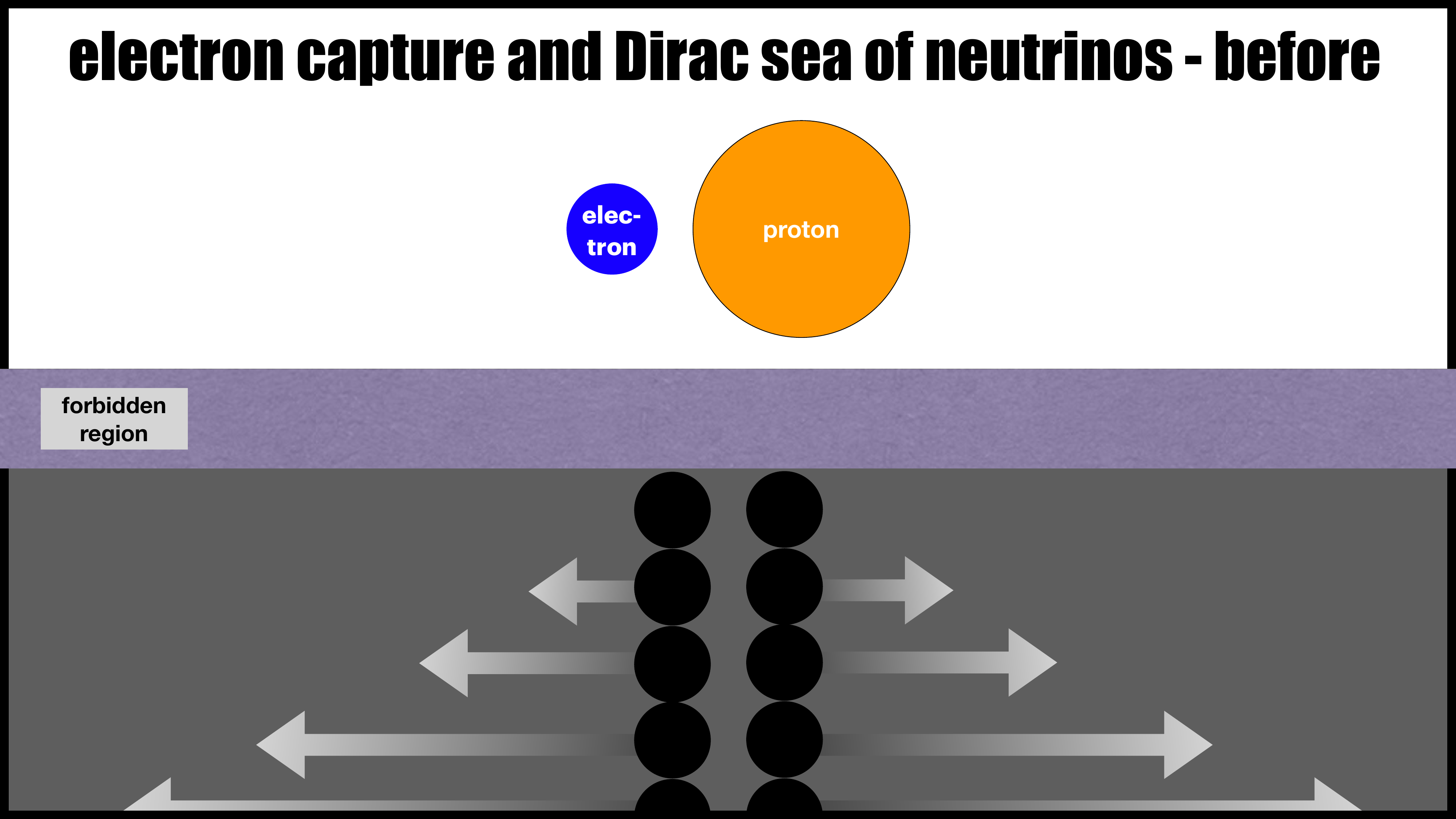}}
\vskip1mm
\centerline{\includegraphics[width=0.99\textwidth]{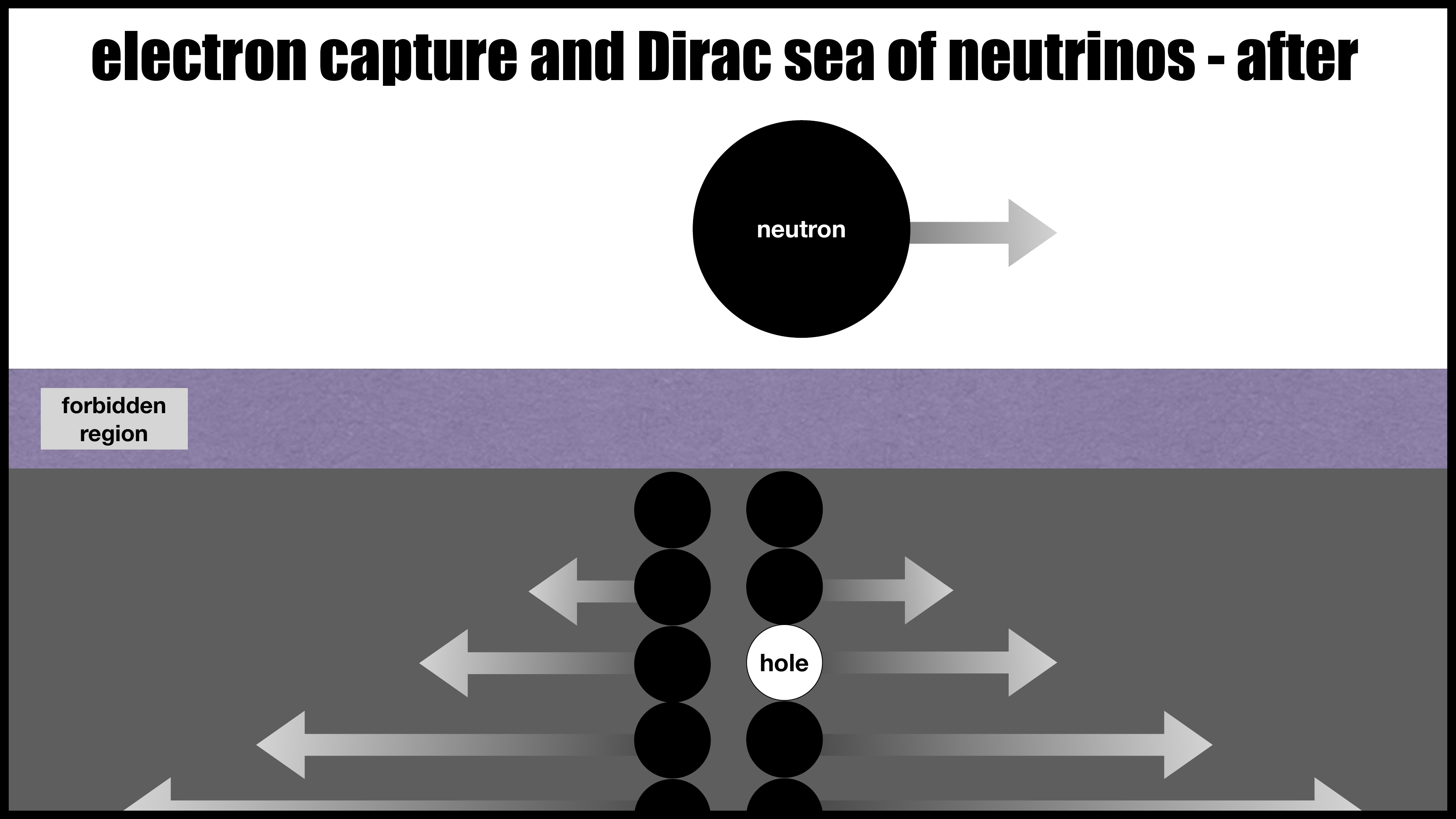}}
\caption{\small\it Description of  electron-proton capture
 in the formalism of second quantization \cite{wick}, emphasizing the 
 Dirac sea of neutrinos (=region of negative energies).  
{\em Panel above:} Initial state of the process; an electron and a proton at rest can be seen; the neutrinos states of Dirac's sea
are all occupied. {\em Panel below:}
Final state of the process. The nucleon changed its isospin state and became a neutron; a hole has formed in the Dirac sea, which can be thought of as an antineutrino, moving in the opposite direction of the neutron (=Dirac hole theory).\label{fig1}}
\end{figure*}

%
%\begin{figure*}[p!]
%\centerline{\includegraphics[width=1\textwidth]{beta1s.pdf}}
%\vskip1mm
%\centerline{\includegraphics[width=1\textwidth]{beta2s.pdf}}
%\caption{\sf\small Description of the neutrino-proton collision according to the
%formalism of the second quantisation. {\em Panel above:} Initial state of the process; one sees a free neutrino, a proton at rest and the Dirac sea of electrons (=region of negative energies) completely occupied.  {\em Panel below:}
%Final state of the process. The nucleon has changed its isospin state, becoming a neutron, and 
%one empty slot in the Dirac sea was formed, which - according to Dirac hole theory - can be thought of as a positron.}
%\end{figure*}

To convince oneself of this, one need only recall three important works  from various parts of the world  \cite{wick,bp,yuk} 
inspired by the `Tentativo' and written soon afterwards, in 1934, by Wick (4 March); Bethe and Peierls (7 April); Yukawa (17 November):
\begin{enumerate}
\item
{\em Wick} derives the predictions for $\beta^+$ emission and electron capture, using - just like Fermi - the second quantisation formalism. The first process explains observations already obtained by Joliot and Curie, the second (one of the proofs for the existence of the neutrino) will receive experimental confirmation a few years later. See Fig.~\ref{fig1} for a description of the latter process\footnote{A description in words is as follows: 
consider Fermi's reaction $n\to p+e +\nu$ taking place on the Dirac sea of the neutrino, but 
supposing that there is a hole that we indicate with  $|\mbox{sea}-\nu\rangle$ (to distinguish it from 
the case when the sea is full). Therefore, the initial state contains a neutron and a anti-neutrino $\bar\nu$, and the final one a proton, an electron, a neutrino and a anti-neutrino.
However, the newly produced neutrino can fill the hole,  and we conclude that the transition $n +\bar{\nu} \to p+e$ is possible. The last step is simply to invert the direction of the arrow getting: $ p+e\to n+\bar{\nu}$.}.
\item {\em Bethe and Peierls}, making explicit reference to Fermi 1933, estimate the neutrino-nucleon interaction cross section by means of a brilliant argument. This reaction will be exploited for the first experimental observation of the neutrino. 
See \cite{nat}.
\item{\em Yukawa}, interested in understanding interactions between nucleons, will propose the idea that interactions between nucleons and those between leptons are mediated by a boson with non-zero mass, in order to reproduce Fermi's theory by mimicking even more closely the structure of electromagnetic interactions.
\end{enumerate}
%
%Per convincersene, basta ricordare tre importanti lavori \cite{wick,bp,yuk} 
%ispirati
%proprio dal `Tentativo' e scritti subito dopo, nel 1934, da Wick (4
%Marzo); Bethe e Peierls (7 Aprile); Yukawa (17 Novembre).
%\begin{enumerate}
%\item
%Wick deriva le predizioni per l'emissione \(\beta^{+}\) e per la cattura
%elettronica, utilizzando - proprio come Fermi - il formalismo di seconda
%quantizzazione. Il primo processo spiega delle osservazioni gi\`a ottenute
%da Joliot e Curie, il secondo (una delle prove dell'esistenza del
%neutrino) ricever\`a conferma sperimentale da l\`i a pochi anni.
%\item
%Bethe e Peierls, facendo esplicito riferimento al lavoro del 1933,
%stimeranno la sezione d'urto di interazione neutrino-nucleone, che
%permetter\`a la prima rivelazione sperimentale. Si vedano le figure 1 e 2
%per la descrizione di questo processo nel contesto della teoria di
%Dirac, Jordan, Klein, o di seconda quantizzazione - quella adottata da
%Fermi.
%\item 
%Yukawa, interessato a capire le interazioni tra nucleoni, proporr\`a
%l'idea, ispirata dalla natura delle interazioni elettromagnetiche e
%basata sulla teoria di Fermi, che le interazioni tra nucleoni e quelle
%tra leptoni siano mediate da un bosone dotato di massa non nulla.
%\end{enumerate}

\subsection{Subsequent progresses of $\beta$ decay theory}

Nowadays, most particle physicists are aware of certain results of the theory of weak interactions, due to subsequent 
theoretical developments. For example, it is generally recalled  that Gamow and
Teller's \cite{gt} included the effect of spin in the  nucleonic current, 
which using the current language of $\gamma$ matrices \cite{pauli36} 
we attribute to the presence of axial currents; even  better known is the
much later history of how the   \(V - A\) structure  (chiral interactions)  of the charged currents 
was understood - see e.g., the fine work of
review by Weinberg~\cite{wei}. 
Among the other recent developments we mention
at least the understanding of the conservation of leptonic number in the
$\beta$ interactions, and the thorough examination of the structure of currents
concluded and completed with the Cabibbo theory. All these advances
dovetail with Fermi's theory.

%Al giorno d'oggi, buona parte dei fisici delle particelle hanno presenti
%certi risultati della teoria delle interazioni deboli, dovuti a sviluppi
%teorici successivi al lavoro di Fermi. Per esempio, si ricordano Gamow e
%Teller (1938 \& 1939) \cite{gt}, che includono l'effetto dello spin nella corrente
%tra nucleoni, che col linguaggio delle matrici gamma \cite{pauli36} (Pauli 1936) attribuiamo all'effetto di correnti assiali; \`e ancora pi\`u nota la
%storia molto successiva di come si sia capita la struttura \(V - A\) ad
%essa collegata (interazioni chirali) - vedi p.e.~il bel lavoro di
%rassegna di Weinberg~del 2009 \cite{wei}. Tra le evoluzioni pi\`u tarde menzioniamo
%almeno la comprensione della conservazione del numero leptonico nelle
%interazioni beta, e l'esame accurato della struttura delle correnti
%conclusosi e completatosi con la teoria di Cabibbo. Sono tutti progressi
%che si innestano e si armonizzano con la teoria di Fermi.

%\begin{figure}[b!]
%
%\caption{\sf\small Description of the neutrino-proton collision according to the
%formalism of the second quantisation. Final state of the process. The nucleon has changed its isospin state, becoming a neutron, and 
%one hole in the Dirac sea was formed, which can be thought of as a positron.
%Compare with Fig.~\ref{fig1}.\label{fig2}}
%\end{figure}

We would just like to point out an important advance that occurred in the late 1930s, which is not sufficiently appreciated today but is the source of many other advances. 
We refer to  a different procedure of quantization of fermionic fields introduced by another of the boys from via Panisperna, Ettore Majorana~\cite{maj}. 
Apart from a witty choice of basis for $\gamma$ matrices used, the new procedure of
quantisation of fermions is exactly the one used today, i.e., the
`canonical quantisation'.  
In the first part of the summary of his work of 1937 (the last one) \cite{maj}, at page 171, we read 
\begin{quote}
\small\sf
It is shown how to achieve a full
formal symmetrization of the quantum theory
of the electron and positron by making use of a new
quantisation process. The meaning of the equations of DIRAC
equations is quite modified and there is no longer any need to speak of states of negative energy.
\end{quote}
To ascertain Fermi's appreciation of
this result, let us read his judgement for the chair competition,
held in the same year (from \cite{frumo}, preface, page xiii):
\begin{quote}
\small\sf
[Majorana] devised a brilliant method for treating the positive and negative electron symmetrically, finally eliminating the need to resort to the extremely artificial and unsatisfactory hypothesis of an infinitely large electric charge spread throughout space, an issue that had been addressed in vain by many other scholars  
\end{quote}
If the terms were used literally, only from this moment on would it be legitimate to speak of a ``vacuum state'' rather than a ``ground state''. In more, evocative terms we can say that it was Majorana who showed the world {\em how to empty the Dirac sea}\footnote{It should be stressed that the Dirac sea hypothesis, unattractive from a physical point of view and now abandoned, is accompanied by relatively simple and almost spontaneous expressions for the second quantization fields, Eq.~\ref{cicciolo}. For this reason it maintains a certain interest in learning paths: it allows us to appreciate how we arrived at modern quantized field theory.}. But an unsuspecting reader, who believed Amaldi's 1984 notations to be the original ones, would not even notice this breakthrough; losing sight of the context, he would no longer be able to truly understand Fermi's work, let alone Majorana's.
 
%In evocative terms: it was Majorana who showed the world how to empty Dirac  sea. Also this progress, 
%that a modern but unaware reader would give for granted from the notations adopted in~\cite{amuld}, ultimately contributes to the corroboration of Fermi's %theory.

\section{Pauli, Fermi and Majorana: three ideas on the neutrino compared}

In this last section, we address one last conceptual point, and discuss the three different ideas of the neutrino that were formulated in the 1930s:
\begin{enumerate}
\item {\em Pauli 1930} \cite{pauli}
introduced the neutrino as a constituent of the atomic nucleus in 1930 and assumed that this particle is emitted in $\beta$ decay. 
This model has no relativistic characteristics and in particular has no connection with Dirac  idea of antimatter.
\item {\em Fermi 1933-1934} \cite{rs33,nc34,zs34}, on the other hand, describes neutrinos that are 
relativistic fermions, completely 
analogous to the electron. Given the formalism adopted - which requires a Dirac sea of neutrinos with negative energy - antineutrinos exist and are quite distinct from neutrinos: see again Fig.~\ref{fig1}.
(In other words, such a neutrino concept corresponds closely to what is now called the `Dirac neutrino'. Although this term is widespread today, Fermi does not use it and there is no work by Dirac describing such a neutrino concept.)
\item Finally, {\em Majorana 1937} neutrino idea \cite{maj} is still different, and consists of the assumption that the neutrino and the antineutrino are the same particle. A similar identification applies for example to the photon, which however, unlike the neutrino, is not a particle of matter.

\end{enumerate}

\noindent Here is how Majorana concludes the summary of his work (again in~\cite{maj}, p.171)
\begin{quote}
\small\sf there is no longer reason [...] to assume for any other type of particles, particularly neutral ones, the existence of~\guillemotleft antiparticles\guillemotright\ corresponding to~\guillemotleft vacua\guillemotright\ of negative energy. 
\end{quote}
where we note the statement on neutral particles which makes implicit reference to neutrinos.
(The reference is made explicit in the text.)

\begin{figure}[t!]
\centerline{\includegraphics[width=0.99\textwidth]{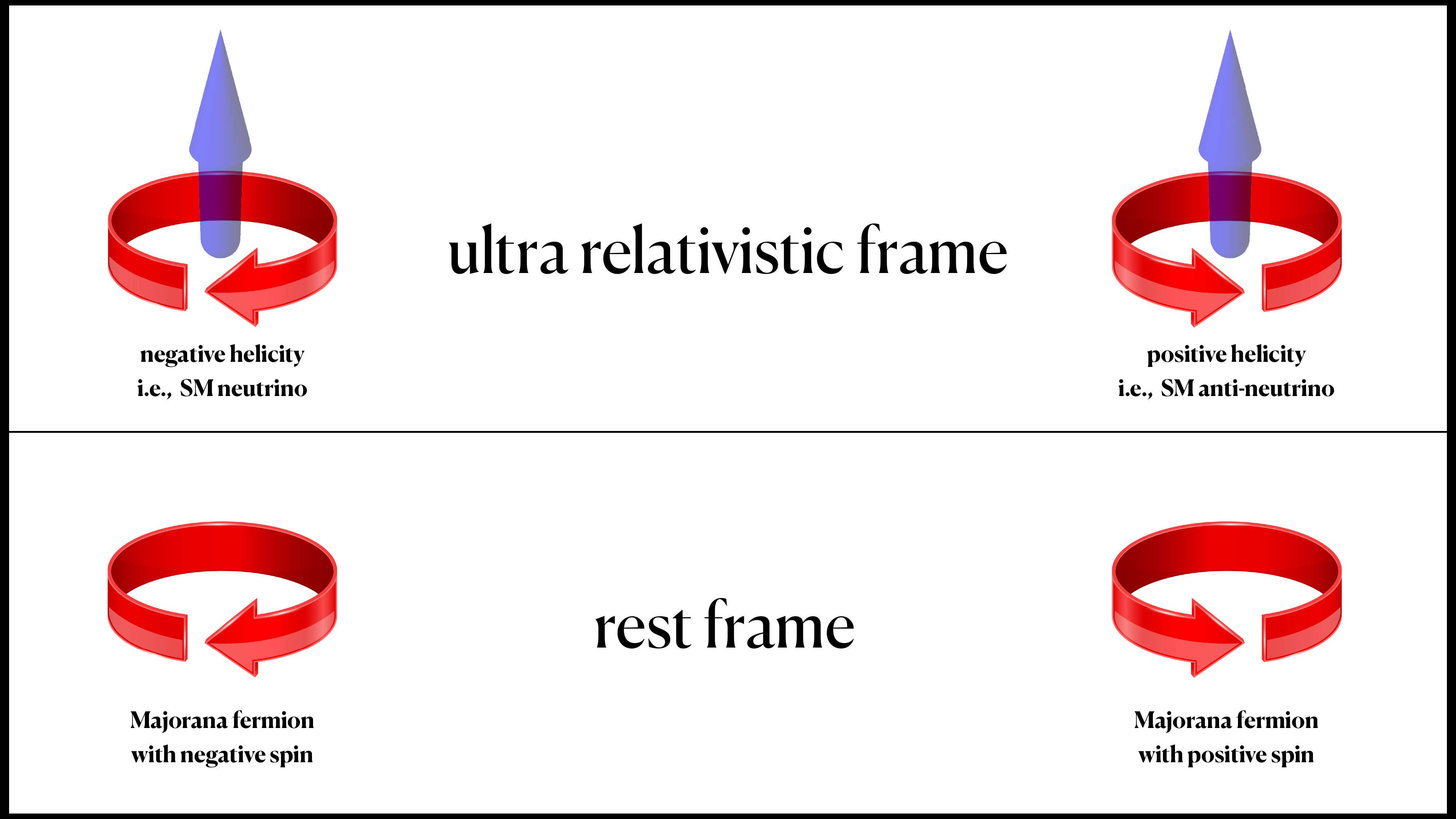}}
\caption{\small\it Illustration of the concept of (neutrino with)
Majorana mass  in the context of the electroweak/\emph{V-A} theory/standard model. The projection of spin onto the momentum of the
particle - helicity - makes it possible to univocally tell 
 neutrinos from antineutrinos in the  ultra-relativistic limit. But in the rest system - which for massive 
fermions exists - the two states are identical, up to 
the orientation of the spin.\label{fig3}}
\end{figure}

We conclude by remarking that the structure of the ``standard model'' of the
electroweak interactions - and in particular, the chiral nature of the charged-currents
weak interactions and the way in which neutrinos are included -
suggests that Majorana's hypothesis is realised in nature, albeit in a quite specific way: 
the neutrino and the antineutrino, which we know to be
different from each other when they move in  ultra-relativistic motion, 
manifest themselves as \emph{the same particle in the system at rest}. 
Fig.~\ref{fig3} better illustrates the physical content of this
statement. This hypothesis on the nature of the neutrino 
is the subject of lively experimental investigations in laboratories all over the world\footnote{For two recent reviews on this subject, see \cite{mois} and \cite{tois}. For an interesting discussion of the influence of Hermann Weyl's ideas in neutrino physics, see \cite{bia}.}.
  
  For a more detailed discussion and further references, we refer 
the reader to \cite{90}; for the modern developments of 
Fermi's theory, see \cite{barb}.

%
%Concludiamo osservando che la struttura del modello standard delle
%interazioni elettrodeboli - ed in particolare, la natura chirale delle
%interazioni deboli ed il modo in cui sono inclusi i neutrini -
%suggerisce che l'ipotesi di Majorana sia realizzata in natura, seppure in un modo
%abbastanza specifico: il neutrino e l'antineutrino, che sappiamo essere
%diversi tra di loro quando si muovono di moto  ultra-relativistico, si
%manifestano essere la stessa particella \emph{nel sistema a riposo}. La
%figura 3 illustra meglio il contenuto fisico di questa
%affermazione\footnote{Majorana's hypothesis is the subject of lively experimental investigations in laboratories all over the world. For two recent reviews on this subject, see \cite{mois} and \cite{tois}.}.
%
%  Per una discussione pi\`u articolata ed ulteriori riferimenti, rimandiamo
%il lettore a \cite{90}; per le evoluzioni moderne della teoria di
%Fermi, si veda \cite{barb}.

\paragraph{Acknowledgments}

{\small
I thank Salvatore Esposito for the precious discussion and Luigi Romano for attentive reading.
Work partially supported by grant
{\em PANTHEON: Perspectives in Astroparticle and Neutrino
THEory with Old and New Messengers} no.\ 2022E2J4RK,   
part of 
PRIN 2022 programme funded by the Ministry of 
University and Research (MUR). 

}

%Questo lavoro \`e stato parzialmente sostenuto dalla borsa di ricerca
%numero 2022E2J4RK "PANTHEON: Perspectives in Astroparticle and Neutrino
%THEory with Old and New messengers" nell'ambito del
%programma PRIN 2022 finanziato dal Ministero
%dell'Universit\`a e della Ricerca (MUR).

%
%\vfill
%
%\parskip1ex
%\tableofcontents
\end{document}